\documentclass[aps,english,showpacs,twocolumn,pra]{revtex4-1}
\usepackage{amssymb}
\usepackage{amsmath}
\usepackage{graphicx}
\usepackage{epsfig}

\begin{document}

\title{Parity-time symmetry under magnetic flux}
\author{L. Jin}
\email{jinliang@nankai.edu.cn}
\author{Z. Song}
\affiliation{School of Physics, Nankai University, Tianjin 300071, China}

\begin{abstract}
{We study a parity-time ($\mathcal{PT}$) symmetric ring lattice, with one
pair of balanced gain and loss located at opposite positions. The system
remains $\mathcal{PT}$-symmetric when threaded by a magnetic flux; however,
the $\mathcal{PT}$ symmetry is sensitive to the magnetic flux, in the
presence of a large balanced gain and loss, or in a large system. We find a
threshold gain/loss, above which, any nontrivial magnetic flux breaks the $%
\mathcal{PT}$ symmetry. We obtain the maximally tolerable magnetic flux for
the exact $\mathcal{PT}$-symmetric phase, which is approximately linearly
dependent on a weak gain/loss.}
\end{abstract}

\maketitle


\section{Introduction}

A magnetic flux enclosed in electron trajectories induces electron wave
function interference; this is the well-known magnetic Aharonov-Bohm (AB)
effect~\cite{ES,AB}. The AB effect describes a quantum phenomenon in which a
charged particle is affected by the vector potential of an electromagnetic
field. The AB effect has inspired breakthroughs in modern physics. Photons
are known to be neutral particles that do not directly interact with
magnetic fields. However, the magnetic AB effect for photons has been
proposed and realized by magnetic-optical effects~\cite{MOeffect}, dynamical
modulation~\cite{DynamicModulation}, and photon-phonon interaction~\cite%
{PhotPhot}. This is attributed to an effective magnetic field originating
from the fictitious gauge field felt by photons, where photons behave as
electrons in a magnetic field. The concept of effective magnetic field for
photons provides new opportunities in optics; it stimulates interest in the
exploration of fundamental physics and in the creation of applications~\cite%
{MOeffect,DynamicModulation,PhotPhot,Hafezi2011,Hafezi2013,Chong}. In
particular, parity-time ($\mathcal{PT}$) symmetric optical systems proposed
with effective magnetic flux allow nonreciprocal light transport~\cite%
{Longhi2015OL,LXQ}.

The $\mathcal{PT}$-symmetry has attracted tremendous interest over the last
decade. A $\mathcal{PT}$-symmetric system may possess an entirely real
spectrum even though it is non-Hermitian~\cite%
{Bender98,Znojil99,Dorey01,Ali02}. The $\mathcal{PT}$-symmetric system is
invariant under the combined parity ($\mathcal{P}$) and time-reversal ($%
\mathcal{T}$) operators, and its potential fulfills $V^{\ast }\left(
x\right) =V\left( -x\right) $. In 2007, a coupled optical waveguide system
with an engineered refractive index and gain/loss profile was proposed to
realize a $\mathcal{PT}$-symmetric structure~\cite{Musslimani OL}. The
proposal was based on a classical analogy, namely, that Maxwell's equations,
describing light propagation under paraxial approximation, are formally
equal to a Schr\"{o}dinger equation~\cite{QuanOptAnology}. Thereafter, a
number of intriguing phenomena can be predicted and experimentally verified,
either in a passive or in an active $\mathcal{PT}$-symmetric optical
structure, including parity-time ($\mathcal{PT}$) symmetry breaking~\cite%
{AGuo}, power oscillation~\cite{CERuter}, coherent perfect absorbers~\cite%
{CPA,HChen}, spectral singularities~\cite{AliPRL2009}, unidirectional
invisibility~\cite{Lin}, and nonreciprocal wave propagation~\cite{Feng}.
Recently, an active $\mathcal{PT}$-symmetric optical system has been
realized using two coupled microcavities~\cite{PengNP}, and optical gain
played a key role. Optical isolators~\cite{PengNP,PengNC,Chang} and $%
\mathcal{PT}$-symmetric lasers~\cite{Jing,PengScience,FengZhang,Hodaei} were
demonstrated; the gain induced a large optical nonlinearity. Both the
coupled optical waveguides and microcavities were described by a
tight-binding model. The tight-binding model demonstrated analytical and
numerical tractability for study of the $\mathcal{PT}$ symmetry~\cite%
{Longhi1,Longhi2,Znojil1,Znojil2,Jin,Bendix}. The $\mathcal{PT}$-symmetric
phase diagram, as well as the wave packet dynamics in $\mathcal{PT}$%
-symmetric systems with open boundary conditions, have been investigated~%
\cite{Joglekar1,Joglekar2,Joglekar3,Joglekar4}. Although the properties of a
lattice with open boundary conditions and the properties of a lattice with
periodical boundary conditions are similar as large systems approach a size
limit, the differences are notable when the system size is small~\cite%
{Joglekar2012}. Currently, most experimentally accessible $\mathcal{PT}$%
-symmetric systems are small in size~\cite{CERuter,PengNP,Chang}; hence,
studying $\mathcal{PT}$-symmetric systems under periodical boundary
conditions is worthwhile, and some pioneering works have already focused on
this~\cite{Joglekar2012,JinHu,Joglekar2013,Longhi2013}.

\begin{figure}[tbp]
\includegraphics[ bb=60 440 520 720, width=7.5 cm, clip]{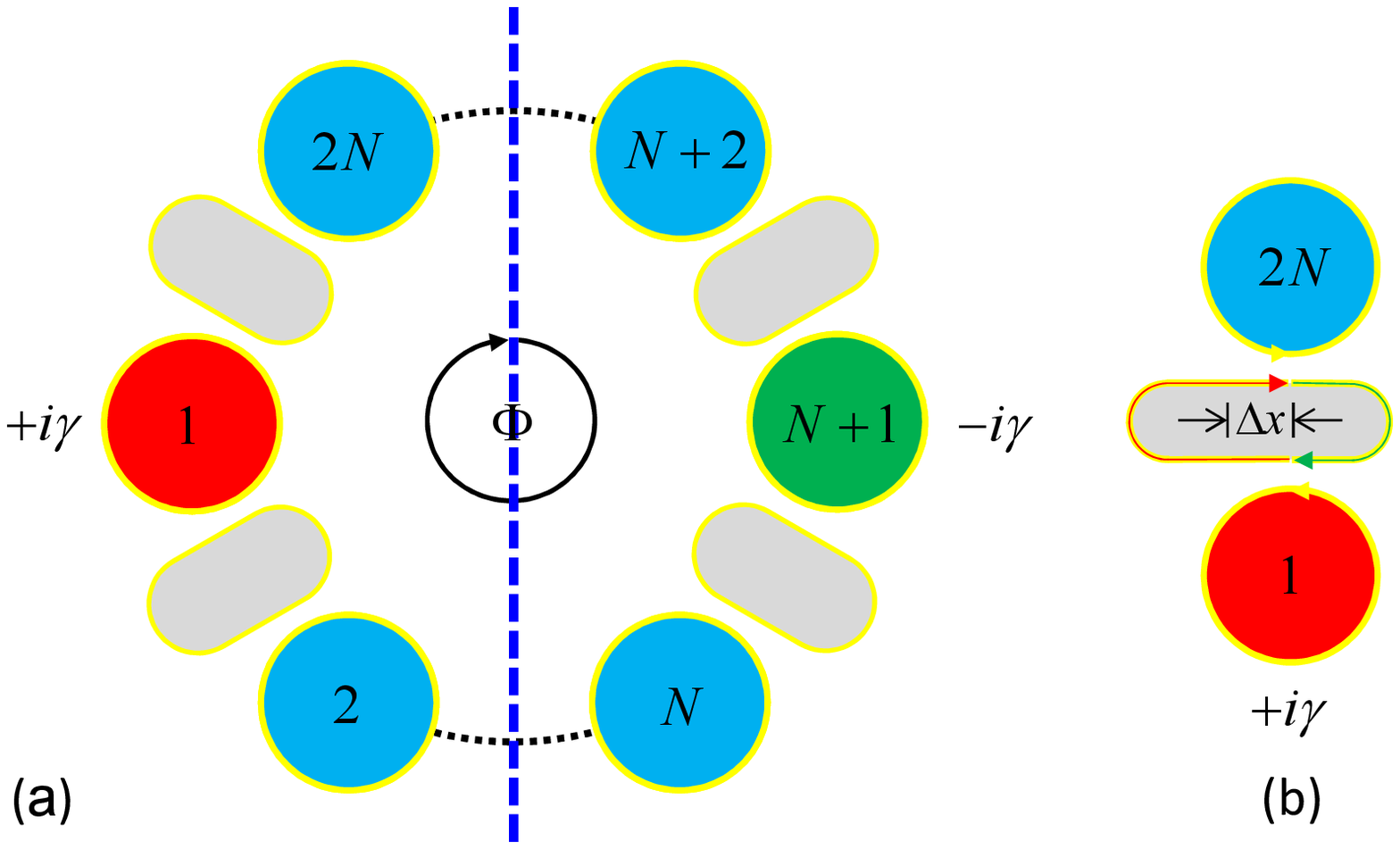}
\caption{(Color online) (a) Schematic illustration of a $\mathcal{PT}$-symmetric lattice model using coupled resonators threaded by a magnetic flux $\Phi$. (b) Effective magnetic flux $\Phi$ is introduced between resonator $1$ and $2N$ through an anti-resonant
auxiliary resonator (grey). The fact that the forward- (green arrow) and backward-going (red arrow) path
lengths difference $\Delta x$, induces a nonreciprocal hopping phase $e^{\pm i 2\pi \Delta x/\lambda}$.
} \label{fig1}
\end{figure}

In this paper, we study the influence of magnetic flux on the $\mathcal{PT}$
symmetry in coupled resonators in ring configuration (Fig.~\ref{fig1}a). The
system is modeled by a magnetic tight-binding lattice, with a Peierls phase
factor in the hopping between adjacent sites. The one dimensional ring
system has a single pair of balanced gain and loss located at opposite
positions; when threaded by magnetic flux, the ring system remains $\mathcal{%
PT}$-symmetric. We reveal that the $\mathcal{PT}$ symmetry is sensitive to
the enclosed magnetic flux, especially when the gain/loss or system size is
large. We find a threshold for the gain/loss, above which, any nontrivial
magnetic flux breaks the $\mathcal{PT}$ symmetry. We also enumerate the
eigenvalues that have broken $\mathcal{PT}$ symmetry. The threshold
gain/loss, as well as the $\mathcal{PT}$-symmetry breaking levels, are
tunable by the magnetic flux. We present a $\mathcal{PT}$-symmetric phase
diagram in the gain/loss and magnetic flux parameter spaces. The maximally
tolerable magnetic flux for the exact $\mathcal{PT}$-symmetric phase is
found for a $2N $-site ring system. Our findings offer insights regarding
magnetic flux in $\mathcal{PT}$-symmetric systems and might provide useful
applications of $\mathcal{PT} $ symmetry in quantum metrology.

This paper is organized as follows. We formulate a $\mathcal{PT}$-symmetric
Hamiltonian threaded by a magnetic flux in Sec.~\ref{II}. We show the energy
spectrum and phase diagram of a four-site ring system in Sec.~\ref{III}. We
elucidate in detail that the $\mathcal{PT}$ symmetry is significantly
affected by the enclosed magnetic flux through studying a $2N$-site ring
system in Sec.~\ref{IV}. Our conclusions are drawn in Sec.~\ref{V}.

\section{Model and formalism}

\label{II}

We consider a discrete $\mathcal{PT}$-symmetric ring system threaded by a
magnetic flux. The system is schematically illustrated in Fig.~\ref{fig1}a
using coupled resonators, which constitute a $2N$-site lattice system
described by a tight-binding model. The enclosed magnetic flux in the ring
system is denoted as $\Phi $. The magnetic flux acts globally: circling
photons or charged particles accumulate a phase factor $e^{\pm i\Phi }$ in
each round because of the AB effect. The plus/minus sign represents a
clockwise/counterclockwise direction of motion. The magnetic flux affects
the system periodically; the period is $2\pi $, which corresponds to one
quantum of the effective magnetic flux. A balanced pair of gain and loss is
symmetrically located at opposite positions on sites $1$ and $N+1$,
respectively. The Hamiltonian for this $2N$-site ring system is given by:
\begin{equation}
H_{\mathrm{flux}}=-\sum_{j=1}^{2N}(e^{i\phi }a_{j}^{\dagger }a_{j+1}+\mathrm{%
h.c.})+i\gamma (a_{1}^{\dagger }a_{1}-a_{N+1}^{\dagger }a_{N+1}),
\end{equation}%
where $a_{j}^{\dagger }$ ($a_{j}$) is the creation (annihilation) operator
for site $j$, the periodic boundary condition requires $a_{2N+j}^{\dagger
}=a_{j}^{\dagger }$. The hopping strength between adjacent sites is set to
unity without loss of generality. The balanced gain and loss are conjugate
imaginary potentials on sites $1$ and $N+1$, the rate is $\gamma $ ($\gamma
>0$). The enclosed magnetic field effectively induces an additional phase
factor $e^{\pm i\phi }$ in the hoppings (Fig.~\ref{fig1}b), where $\phi
=\Phi /\left( 2N\right) $ is an averaged additional phase between adjacent
sites.

The parity operator $\mathcal{P}$ is defined as $\mathcal{P}j\mathcal{P}%
^{-1}\rightarrow N+2-j$, the time-reversal operator $\mathcal{T}$ is defined
as $\mathcal{T}i\mathcal{T}^{-1}\rightarrow -i$. We note that in the
presence of a nontrivial phase factor ($e^{i\Phi }\neq 1$), the system
Hamiltonian $H_{\mathrm{flux}}$ remains invariant under the combined $%
\mathcal{PT}$ operator, i.e., $(\mathcal{PT})H_{\mathrm{flux}}(\mathcal{PT}%
)^{-1}=H_{\mathrm{flux}}$. The ring system is $\mathcal{PT}$-symmetric with
respect to its central axis (dashed blue line shown in Fig.~\ref{fig1}a).
When $H_{\mathrm{flux}}$ is in its exact $\mathcal{PT}$-symmetric phase, its
spectrum is entirely real and all eigenstates are $\mathcal{PT}$-symmetric.
When $H_{\mathrm{flux}}$ is in its broken $\mathcal{PT}$-symmetric phase,
complex conjugate pairs emerge and corresponding eigenstates are no longer $%
\mathcal{PT}$-symmetric. In the following, we discuss how the exact $%
\mathcal{PT}$-symmetric phase is affected by the parameters of the ring
system, in particular, the enclosed magnetic flux and gain/loss.

\section{Energy spectrum and phase diagram of a four-site ring system}

\label{III} We first consider the simplest case: A four-site $\mathcal{PT} $%
-symmetric ring system with $N=2$. The Hamiltonian for a four-site ring
system is written in the form:
\begin{equation}
H_{\mathrm{flux}}^{\mathrm{[4]}}=-\sum_{i=1}^{4}(e^{i\phi }a_{i}^{\dagger
}a_{i+1}+\mathrm{h.c.})+i\gamma (a_{1}^{\dagger }a_{1}-a_{3}^{\dagger
}a_{3}),
\end{equation}%
where the periodical boundary condition requires $a_{j+4}^{\dagger
}=a_{j}^{\dagger }$. The phase factor $e^{\pm i\phi }$ in front of the
hoppings between adjacent sites indicates that the enclosed magnetic flux in
the ring system is equal to $\Phi =4\phi $. We can diagonalize the $4\times
4 $ matrix to acquire the spectrum of the ring system. The eigenvalue $E$
satisfies $E^{2}\left( 4-E^{2}-\gamma ^{2}\right) =4\sin ^{2}(\Phi /2)$.
Solving the equation, we obtain four eigenvalues
\begin{equation}
E=\pm \sqrt{2}\sqrt{\left( 1-\frac{\gamma ^{2}}{4}\right) \pm \sqrt{\left( 1-%
\frac{\gamma ^{2}}{4}\right) ^{2}-\sin ^{2}\left( \frac{\Phi }{2}\right) }}.
\end{equation}

For nontrivial magnetic flux $\Phi \neq 2m\pi $ ($m\in \mathbb{Z}$) changes,
the system spectrum changes as the magnetic flux $\Phi $. At $\gamma =0$,
the energy levels shift and form two pairs of doubly degenerate states with
energy $E=\pm \sqrt{2}$ when $\Phi =2m\pi +\pi $ ($m\in \mathbb{Z}$), where
the $\mathcal{PT}$ symmetry of the eigenstates is extremely sensitive to the
balanced gain and loss, i.e., any nonzero gain/loss ($\gamma \neq 0$) breaks
the $\mathcal{PT}$ symmetry.

\begin{figure}[tb]
\includegraphics[bb=0 0 270 260, width=7.0 cm, clip]{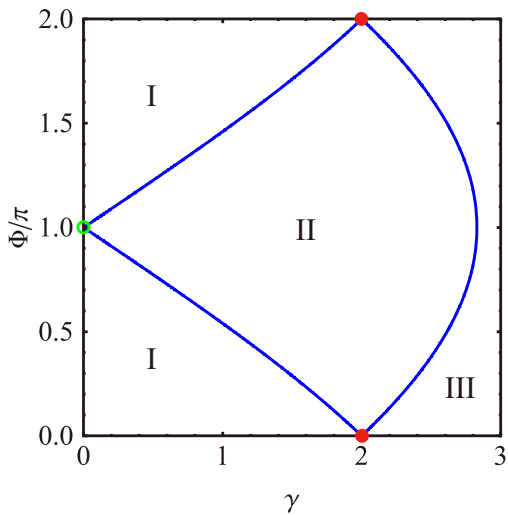}
\caption{(Color online) Phase diagram of a four-site ring system in the parameter spaces $\gamma$
and $\Phi$. Eigenvalues in region I: Real; region II: Complex conjugate pairs; and region III: Pure imaginary conjugate pairs.
Region I is the exact $\mathcal{PT}$-symmetric phase. Region II and III compose the broken $\mathcal{PT}$-symmetric phase.
The boundaries (blue curves) indicate the exceptional points where two eigenstates coalescence occurs. The red dots represent the coalescence of three eigenstates, and the green circle represents a Hermitian system without coalescence of eigenstates.}
\label{region}
\end{figure}

In the general case, a magnetic flux $\Phi $ breaks the $\mathcal{PT}$
symmetry of the eigenstates in the situation that $\cos \Phi <1-2(1-\gamma
^{2}/4)^{2} $. The $\mathcal{PT}$-symmetric phase diagram of a four-site
ring system is shown in Fig.~\ref{region}. Region I represents the exact $%
\mathcal{PT}$-symmetric phase, regions II and III compose the broken $%
\mathcal{PT}$-symmetric phase.

The magnetic flux $\Phi $ plays different roles as the system gain/loss $%
\gamma $ varies in regions of different parameters. For gain/loss $\gamma =0$%
, the dispersion relation is $E=-2\cos (k+\Phi /4)$ with eigenvector $k=0 $,
$\pi /2$, $\pi $, $3\pi /2$. In this case, the magnetic flux $\Phi $ shifts
the energy spectrum with eigenstates unchanged.

Region I: As $\gamma $ increases, the ring system is in its exact $\mathcal{%
PT}$-symmetric phase when $\cos \Phi \geqslant 1-2(1-\gamma ^{2}/4)^{2}$ in $%
0<\gamma <2$. Region II: The $\mathcal{PT}$ symmetry is broken when $\cos
\Phi <1-2(1-\gamma ^{2}/4)^{2}$, eigenvalues that are complex conjugate
pairs appear and the magnetic flux $\Phi $ changes both the energy (real
part of eigenvalues) and the amplification/decay (imaginary part of
eigenvalues) of the eigenstates. In this case, the region with eigenvalues
that are complex conjugate pairs expands in $0<\gamma <2$ but shrinks in $%
\gamma >2$ as $\gamma $ increases. Region III: The magnetic flux $\Phi $
changes only the amplification/decay of the eigenstates when $\cos \Phi
\geqslant 1-2(1-\gamma ^{2}/4)^{2}$ in $\gamma \geqslant 2$. This is because
the eigenvalues are purely imaginary conjugate pairs in this situation.

In the trivial magnetic flux $\Phi =2m\pi $ ($m\in \mathbb{Z}$) case, one
complex conjugate pair becomes a two-fold degenerate state with an
eigenvalue of zero in $\gamma >2$; three energy levels coalesce at $\gamma
=2 $ (red dots) with an eigenvalue of zero, and the $\mathcal{PT}$-symmetric
ring system becomes a nondiagonalizable Hamiltonian including a $3\times 3$
Jordan block. Figure~\ref{region} shows the situation in which nontrivial
magnetic flux satisfies $\cos \Phi =1-2(1-\gamma ^{2}/4)^{2}$ (excluding the green circle and red dots, e.g., $\gamma =2\sqrt{2}$, $\Phi =\pi $); this is shown in Fig.~\ref{region} by the blue curves that serve as the boundaries of
different phases. In this case, two energy levels coalesce, and the $%
\mathcal{PT}$-symmetric ring system is a nondiagonalizable Hamiltonian
composed of two $2\times 2$ Jordan blocks.

\section{The effects of magnetic flux in the $\mathcal{PT}$-symmetric $2N$%
-site ring system}

\label{IV} The magnetic flux is gauge-invariant and acts globally in the
ring system. Taking the local transformation $a_{j}^{\dagger }\rightarrow
e^{i\phi j}a_{j}^{\dagger }$, the magnetic flux is unchanged and the
Hamiltonian $H_{\mathrm{flux}}$ changes into $H_{\mathrm{sc}}$ with a
nonreciprocal coupling between sites $1$ and $2N$. The Hamiltonian $H_{%
\mathrm{sc}}$ is given by
\begin{eqnarray}
H_{\mathrm{sc}} &=&-\sum_{i=1}^{2N-1}(a_{i}^{\dagger }a_{i+1}+\mathrm{h.c.}%
)-e^{-i\Phi }a_{1}^{\dagger }a_{2N}-e^{i\Phi }a_{2N}^{\dagger }a_{1}  \notag
\\
&&+i\gamma (a_{1}^{\dagger }a_{1}-a_{N+1}^{\dagger }a_{N+1}),  \label{Hsc}
\end{eqnarray}%
which can be realized in coupled optical resonators by introducing synthetic
magnetic flux, and a balanced gain and loss in the resonators. The synthetic
magnetic flux is introduced through an optical path imbalance method in the
coupling process~\cite{Hafezi2014}. Consider a system with $2N$ coupled
resonators in a ring configuration. The coupling between resonator $1$ and $%
2N$ is an effective coupling induced by an other auxiliary resonator (Fig.~%
\ref{fig1}b). The nonreciprocal phase factor $e^{\pm i\Phi }$ in the hopping
between resonator $1$ and $2N$ is caused by the optical path length
difference $\Delta x$, which is the difference in between the forward-going
and backward-going optical paths of the auxiliary resonator in the coupling
process. The effective magnetic flux introduced is equal to $\Phi =2\pi
\Delta x/\lambda $ where $\lambda $ is the optical wavelength~\cite%
{Hafezi2014}. Therefore, $\Phi $ is proportional to the path length
difference $\Delta x$, and is tunable through changing the position of
auxiliary resonator.

The eigenvalues of the $\mathcal{PT}$-symmetric ring system $H_{\mathrm{sc}}$
is calculated as follows. We denote the wave function for eigenvalue $E_{k}$
as $f_{k}(j)$. The wave function $f_{k}(j)$ is assumed as a superposition of
forward- and backward-going waves,%
\begin{equation}
f_{k}(j)=\left\{
\begin{array}{c}
A_{k}e^{ikj}+B_{k}e^{-ikj},\left( 1\leqslant j\leqslant N+1\right) \\
C_{k}e^{ikj}+D_{k}e^{-ikj},\left( N+1\leqslant j\leqslant 2N\right)%
\end{array}%
\right. .  \label{WA}
\end{equation}
The Schr\"{o}dinger equation for site $j$ in the ring system (excluding site
$1$, $N+1$, and $2N $) is given as:
\begin{equation}
f_{k}(j-1)+f_{k}(j+1)+E_{k}f_{k}(j)=0,  \label{SchEQ}
\end{equation}%
substituting the wave functions Eq.~(\ref{WA}) into Eq.~(\ref{SchEQ}), we
obtain the eigenvalue $E_{k}=-2\cos k$.

\begin{figure}[tb]
\includegraphics[ bb=0 0 580 250, width=8.8 cm, clip]{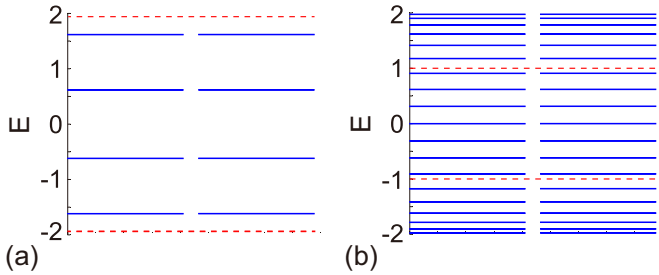}
\caption{(Color online) Energy levels for a ring system under trivial magnetic
flux $\Phi=2m\pi$ ($m\in \mathbb{Z}$) with (a) $N=5$, $\gamma=0.5$; and (b)
$N=20$, $\gamma=\sqrt{3}$. The spectrum includes $N-1$ pairs of two-fold
degenerate energy levels $-2\cos (n\pi /N)$ where $n\in \lbrack 1,N-1]$ (solid blue lines), and also two $\gamma$ dependent energy levels $\pm \sqrt{4-\gamma ^{2}}$
(dashed red lines).} \label{fig_illus}
\end{figure}

The Schr\"{o}dinger equations for sites $1$, $N+1$, and $2N$ are:
\begin{eqnarray}
f_{k}\left( 2\right) +e^{-i\Phi }f_{k}\left( 2N\right) -\left( i\gamma
-E_{k}\right) f_{k}\left( 1\right) &=&0,  \label{SchrodingerEq} \\
e^{i\Phi }f_{k}\left( 1\right) +f_{k}\left( 2N-1\right) +E_{k}f_{k}\left(
2N\right) &=&0, \\
f_{k}\left( N+2\right) +f_{k}\left( N\right) +\left( i\gamma +E_{k}\right)
f_{k}\left( N+1\right) &=&0.
\end{eqnarray}%
From the continuity of wave function on site $N+1$, the wave function $%
f_{k}(N+1)$ should satisfy
\begin{equation}
A_{k}e^{i\left( N+1\right) k}+B_{k}e^{-i\left( N+1\right) k}=C_{k}e^{i\left(
N+1\right) k}+D_{k}e^{-i\left( N+1\right) k},  \label{Continuity}
\end{equation}%
After simplification of the Schr\"{o}dinger equations and the continuity
equation shown in Eqs.~(\ref{SchrodingerEq}-\ref{Continuity}), we derive a
critical equation for eigenvector $k$ as an implicit function of the
gain/loss $\gamma $ and the enclosed magnetic flux $\Phi $. The critical
equation for eigenvector $k$ has the form:
\begin{equation}
\left( 1-\frac{\gamma ^{2}}{4\sin ^{2}k}\right) \sin ^{2}\left( Nk\right)
-\sin ^{2}\left( \frac{\Phi }{2}\right) =0.  \label{critical_Eq}
\end{equation}
In the situation of a trivial magnetic flux $\Phi =2m\pi $ ($m\in \mathbb{Z}$%
), the gain/loss affects only one pair of energy levels; it leaves the
others unchanged. The spectrum of a $2N$-site ring system includes $N-1$
pairs of two-fold degenerate energy levels, i.e., $-2\cos (n\pi /N)$ with $%
n\in \lbrack 1,N-1]$ (indicated by blue lines in Fig.~\ref{fig_illus}); and
one pair of gain/loss dependent energy levels, i.e., $\pm \sqrt{4-\gamma ^{2}}$ (indicated by dashed red lines in Fig.~\ref{fig_illus}). When the gain/loss $\gamma >2$, the ring system is in its broken $\mathcal{PT}$-symmetric phase,
and the system spectrum has one conjugate pair.

In order to analyze the influence of magnetic flux on the system spectrum,
we denote the left side of the critical equation Eq. (\ref{critical_Eq}) as $%
\mathcal{F}\left( \gamma ,\Phi ,k\right) $, which is a function of
parameters $\gamma $, $\Phi $ for eigenvector $k\neq 0$ (note that $k=0$ is
not the eigenvector when $\gamma \neq 0$),
\begin{equation}
\mathcal{F}\left( \gamma ,\Phi ,k\right) =\left( 1-\frac{\gamma ^{2}}{4\sin
^{2}k}\right) \sin ^{2}\left( Nk\right) -\sin ^{2}\left( \frac{\Phi }{2}%
\right) .
\end{equation}

The real eigenvector $k$ with $\mathcal{F}\left( \gamma ,\Phi ,k\right) =0$
corresponds to the real eigenvalue $E_{k}$ of the $\mathcal{PT}$-symmetric $%
2N$-site ring system. For a situation with gain/loss $\gamma >2$, we have $%
\mathcal{F}\left( \gamma ,\Phi ,k\right) <0$ for any nontrivial magnetic
flux $\Phi \neq 2m\pi $ ($m\in \mathbb{Z}$); this indicates that the
spectrum of the ring system is entirely constituted by conjugate pairs
without any real eigenenergy. For magnetic flux $\Phi =2m\pi +\pi $ ($m\in
\mathbb{Z}$), we have $\mathcal{F}\left( \gamma ,\Phi ,k\right) =-\cos
^{2}\left( Nk\right) -\gamma ^{2}\sin ^{2}\left( Nk\right) /(4\sin ^{2}k)<0$%
, the real eigenvector $k$ is absent, i.e., the system spectrum is entirely
constituted by conjugate pairs at $\gamma \neq 0$.

\begin{figure}[t]
\includegraphics[ bb=55 200 460 608, width=7.0 cm, clip]{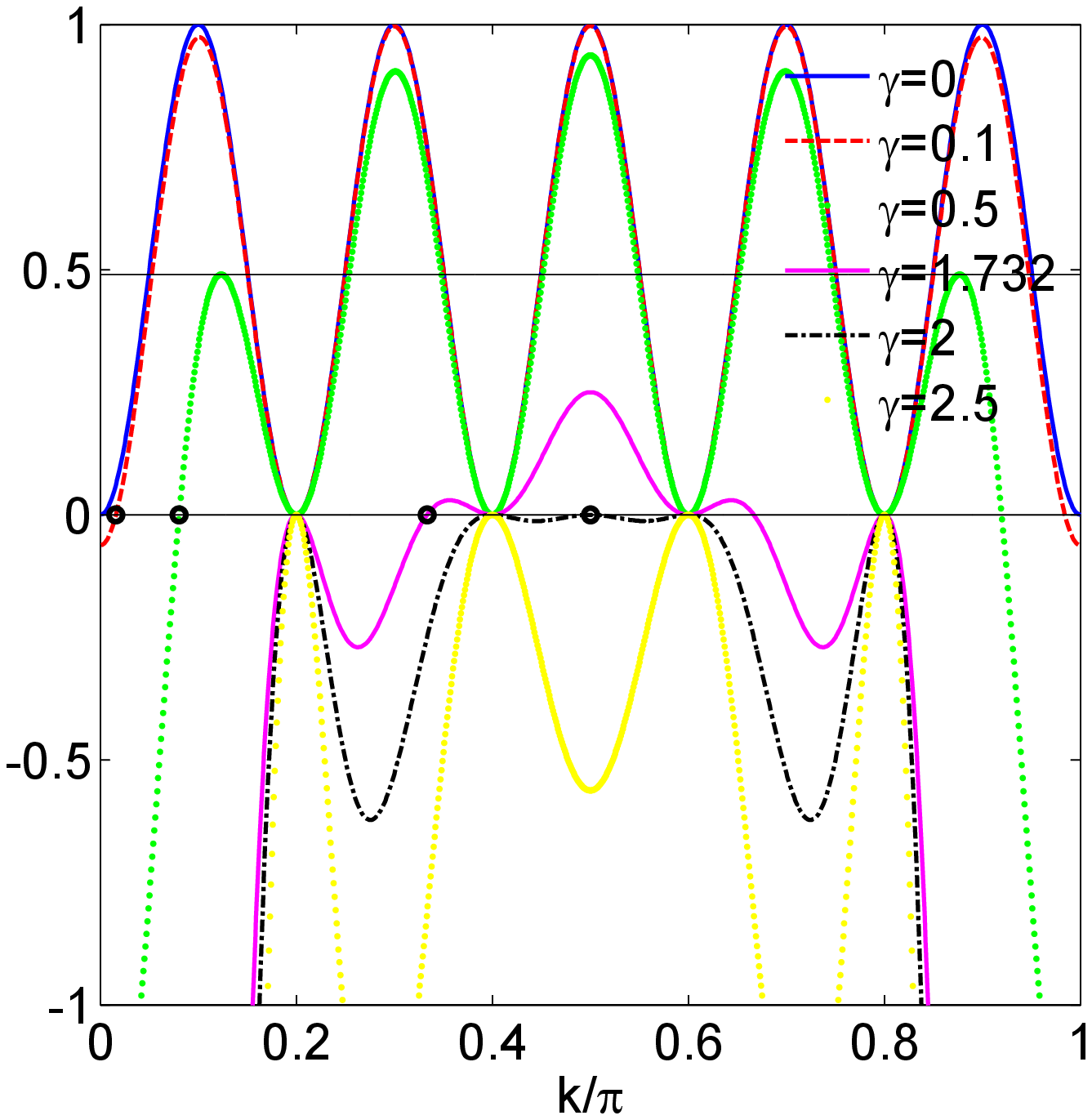}
\caption{(Color online) $\mathcal{F}\left( \protect\gamma ,0,k\right) $ for a ring system with $N=5$ at various gain/loss: $\protect\gamma =0$ (solid blue), $0.1$ (dashed red), $0.5$ (dotted green), $\protect\sqrt{3}$ (solid magenta) , $2$ (dash-dotted black), $2.5$ (dotted yellow). The black circles show $k_{+}=\arccos (\protect\sqrt{1-\protect\gamma ^{2}/4})$ for $\protect\gamma =0.1$, $0.5$, $\protect\sqrt{3}$, $2$. In the
region $\protect\gamma >2$ (e.g. $\protect\gamma =2.5$), no real $k_+$ exists, any $\Phi \neq 2m\pi$ ($m\in\mathbb{Z}$) brings all $2N$ eigenvalues into conjugate pairs. The horizontal black lines are guides to the eye, which indicate $\sin ^{2}{(\Phi/2)}$ for a trivial magnetic flux and the maximal magnetic flux that allows exact $\mathcal{PT}$ symmetry at $\gamma=0.5$.}
\label{F}
\end{figure}

The $\mathcal{PT}$-symmetric $2N$-site ring system has at most one conjugate
pair in its spectrum when magnetic flux is $\Phi =2m\pi $ ($m\in \mathbb{Z}$%
), but has at most $N$ conjugate pairs when magnetic flux is $\Phi =2m\pi
+\pi $ ($m\in \mathbb{Z}$). This implies the system spectrum is sensitive to
magnetic flux, and the number of conjugate pairs appreciably varies with the
magnetic flux. In the following, we systemically investigate how $\mathcal{PT%
}$ symmetry of the eigenstates is affected by the magnetic flux and
gain/loss. We first consider a trivial case of effective magnetic flux with $%
\Phi =2m\pi $ ($m\in \mathbb{Z}$): the function $\mathcal{F}\left( \gamma
,\Phi ,k\right) $ reduces to $\mathcal{F}\left( \gamma ,\Phi ,k\right)
=[1-\gamma ^{2}/(4\sin ^{2}k)]\sin ^{2}\left( Nk\right) $. $\mathcal{F}%
\left( \gamma ,\Phi ,k\right) =0$ is the critical equation for the
eigenvector $k$. We find the eigenvalues are $-2\cos (n\pi /N)$ with $n\in
\lbrack 1,N-1]$ and $\pm \sqrt{4-\gamma ^{2}}$. The gain/loss $\gamma $ only
changes two energy levels, $\epsilon _{\pm }=\mp \sqrt{4-\gamma ^{2}}$, the
eigenvector for $\epsilon _{+}$ is
\begin{equation}
k_{+}=\arccos \sqrt{1-\gamma ^{2}/4}.
\end{equation}%
Hence, when $k_{+}<\pi /N$, the system spectrum may be entirely real even if
$\Phi \neq 2m\pi $ ($m\in \mathbb{Z}$). This indicates a threshold gain/loss
value, $\gamma _{\mathrm{c}}=2\sin \left( \pi /N\right) $. When the
gain/loss is below the threshold ($\gamma <\gamma _{\mathrm{c}}$), the ring
system can be in an exact $\mathcal{PT}$-symmetric phase. As magnetic flux
increases from $0$ to $\pi $, the energy levels with broken $\mathcal{PT}$
symmetry simultaneously emerge from $1$ to $N$ pairs. If $\gamma >\gamma _{%
\mathrm{c}}$, the $\mathcal{PT}$ symmetry is extremely sensitive to the
magnetic flux; any nontrivial magnetic flux $\Phi \neq 2m\pi $ ($m\in
\mathbb{Z}$) breaks the $\mathcal{PT}$ symmetry and causes conjugate pairs
to emerge simultaneously. Above the threshold gain/loss ($\gamma >\gamma _{%
\mathrm{c}}$), the ring system remains in an exact $\mathcal{PT}$-symmetric
phase only when the magnetic flux is $\Phi =2m\pi $ ($m\in \mathbb{Z}$).
Moreover, the minimal number of complex eigenvalue pairs in the presence of
nontrivial magnetic flux is
\begin{equation}
D=2[k_{+}N/\pi ],
\end{equation}%
where $[x]$ stands for the integer part of $x$.

\begin{figure}[t]
\includegraphics[ bb=0 0 360 533, width=8.7 cm, clip]{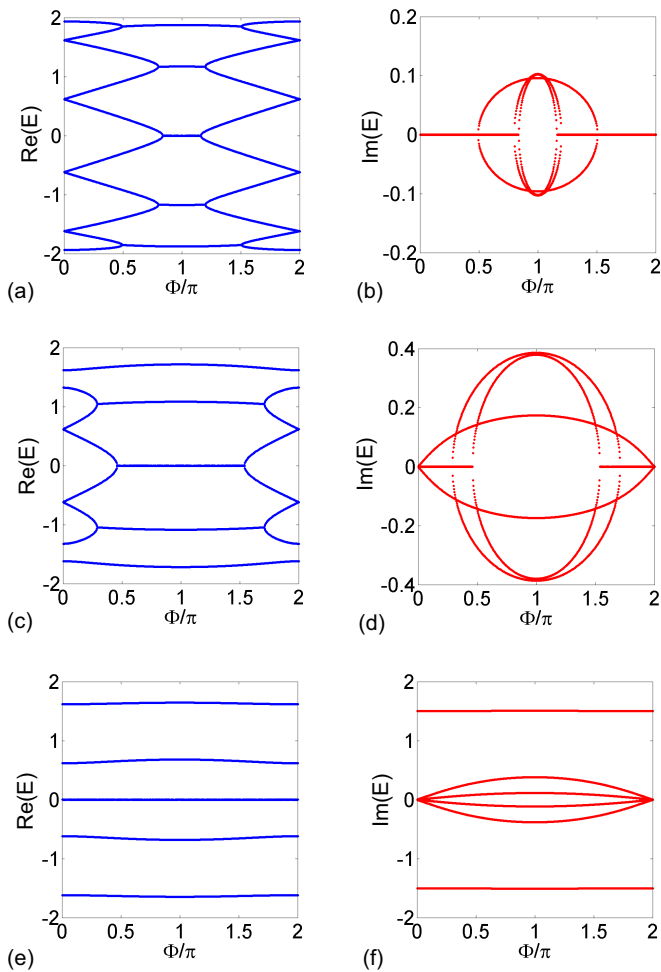}
\caption{(Color online) Energy band of a ring system with $N=5$ at various gain/loss values. (a, b) $\protect\gamma=0.5$, (c, d) $\protect\gamma=1.5$, (e, f) $\protect\gamma=2.5$. The real part (in blue) and imaginary part (in red) are in the left and right panels, respectively.} \label{bands}
\end{figure}

As shown in Fig.~\ref{fig_illus}a, when $\gamma =0.5<\gamma _{\mathrm{c}%
}=2\sin (\pi /5)$, the ring system keeps in the exact $\mathcal{PT}$%
-symmetric phase in the presence of nontrivial magnetic flux. In Fig.~\ref%
{fig_illus}b, $\gamma =\sqrt{3}>\gamma _{\mathrm{c}}=2\sin (\pi /20)$, $%
k_{+}=\pi /3$; the minimal number of energy levels with broken $\mathcal{PT}$
symmetry is $D=12$ for nontrivial magnetic flux; the nontrivial magnetic
flux breaks the $\mathcal{PT}$ symmetry of the energy levels with $|E_{k}|>%
\sqrt{4-\gamma ^{2}}$.

We plot function $\mathcal{F}\left( \gamma ,0,k\right) $ for a ring system
with $N=5 $ in Fig.~\ref{F}. We can see how the eigenvector is changed into
a complex number by the variations of gain/loss $\gamma $ and magnetic flux $%
\Phi $. In Fig.~\ref{F}, $\mathcal{F}\left( \gamma ,0,k\right) $ for ring
system at $\gamma =0$, $0.1$, $0.5$, $\sqrt{3}$, $2$, $2.5$ are plotted. The black circles stand for the eigenvector $k_{+}=\arccos (\sqrt{1-\gamma ^{2}/4%
})$ for $\gamma =0.1$, $0.5$, $\sqrt{3}$, $2$. Note that all $2N$
eigenvalues become conjugate pairs for nontrivial magnetic flux when $\gamma
\geqslant 2$, and for nonzero gain/loss when $\Phi =2m\pi +\pi $ ($m\in
\mathbb{Z}$).

Figure~\ref{bands} shows the energy bands found by numerical diagonalization
of $H_{\mathrm{sc}}$ in Eq. (\ref{Hsc}) with $N=5$; the corresponding
eigenvector $k$ is in accord with the eigenvector from the critical equation
Eq.~(\ref{critical_Eq}). For trivial magnetic flux, the $\mathcal{PT}$
symmetric ring system has only one complex pair when the gain/loss $\gamma
>2 $. For nontrivial flux, the energy level degeneracy (Fig.~\ref{fig_illus}%
) disappears. This is because the nonreciprocal hopping between the coupled
resonators breaks the time-reversal symmetry in the tunneling. The number of
conjugate pairs changes from $1$ ($D$) to $N$ as magnetic flux $\Phi $
increases from $0$ to $\pi $ for gain/loss below (above) the threshold.
Figure~\ref{bands} shows the magnetic flux acting globally, and the spectrum
structure substantially varies with the magnetic flux at different values of
gain/loss.

Now, we focus on the influence of magnetic flux on the $\mathcal{PT}$%
-symmetric phase diagram. In order to get the $\mathcal{PT}$-symmetric phase
diagram, we define $\Phi _{\mathrm{c}}$ as the maximal magnetic flux that
keeps a ring system in its exact $\mathcal{PT}$-symmetric phase. The maximal
magnetic flux $\Phi _{\mathrm{c}} $ depends on the non-Hermitian gain/loss $%
\gamma $ and indicates the boundary of the $\mathcal{PT}$-symmetric phase
diagram in the $\Phi$, $\gamma $ parameter spaces. For a gain/loss above the
threshold, $\gamma >\gamma _{\mathrm{c}}=2\sin \left( \pi /5\right) \approx
1.1756$, any nontrivial magnetic flux $\Phi \neq 2m\pi $ ($m\in \mathbb{Z}$)
breaks the $\mathcal{PT} $ symmetry. In this situation, $\Phi _{\mathrm{c}}$
is zero. The $\mathcal{PT}$ symmetry is extremely sensitive to magnetic flux
for large values of $\gamma $ or $N$ (the gain/loss threshold $\gamma _{%
\mathrm{c}}$ depends on $N$).

For a gain/loss below the threshold, $\gamma <\gamma _{\mathrm{c}}=2\sin
(\pi /5)\approx 1.1756$, e.g. $\gamma =0.5$ (Fig.~\ref{F}, indicated by
dotted green line), the $\mathcal{PT}$ symmetry is robust to the magnetic
flux $\Phi $ when $\sin ^{2}({\Phi /2)}$ is in the region between the two
horizontal black lines shown in Fig.~\ref{F}. In this situation, the
magnetic flux does not break $\mathcal{PT}$ symmetry, and the system
spectrum is entirely real. If $\gamma $ or $\Phi $ increases from $0$, the
system goes through a $\mathcal{PT}$-symmetric breaking phase transmission
when $\mathcal{F}\left( \gamma ,0,k\right) <\sin ^{2}\left( \Phi /2\right) $
for $k\in \lbrack 0,\pi /N]$. Then the entirely real spectrum disappears and
conjugate pairs emerge. For each $\gamma $, the ring system is in its exact $%
\mathcal{PT}$-symmetric phase when the magnetic flux is below the maximal
value $\Phi _{\mathrm{c}}$.

\begin{figure}[t]
\includegraphics[ bb=30 170 500 610, width=4.25 cm, clip]{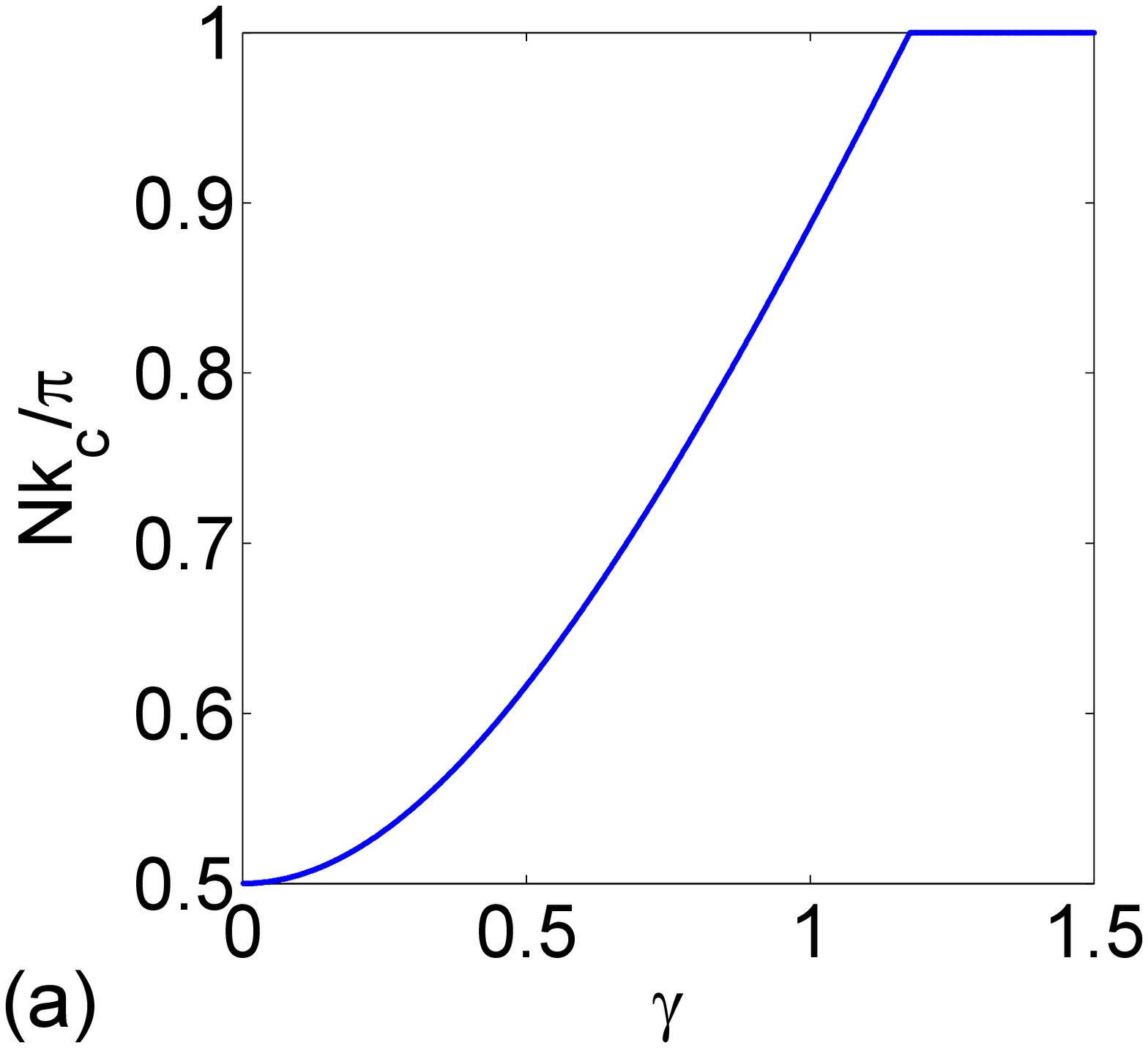} %
\includegraphics[ bb=0 0 470 440, width=4.25 cm, clip]{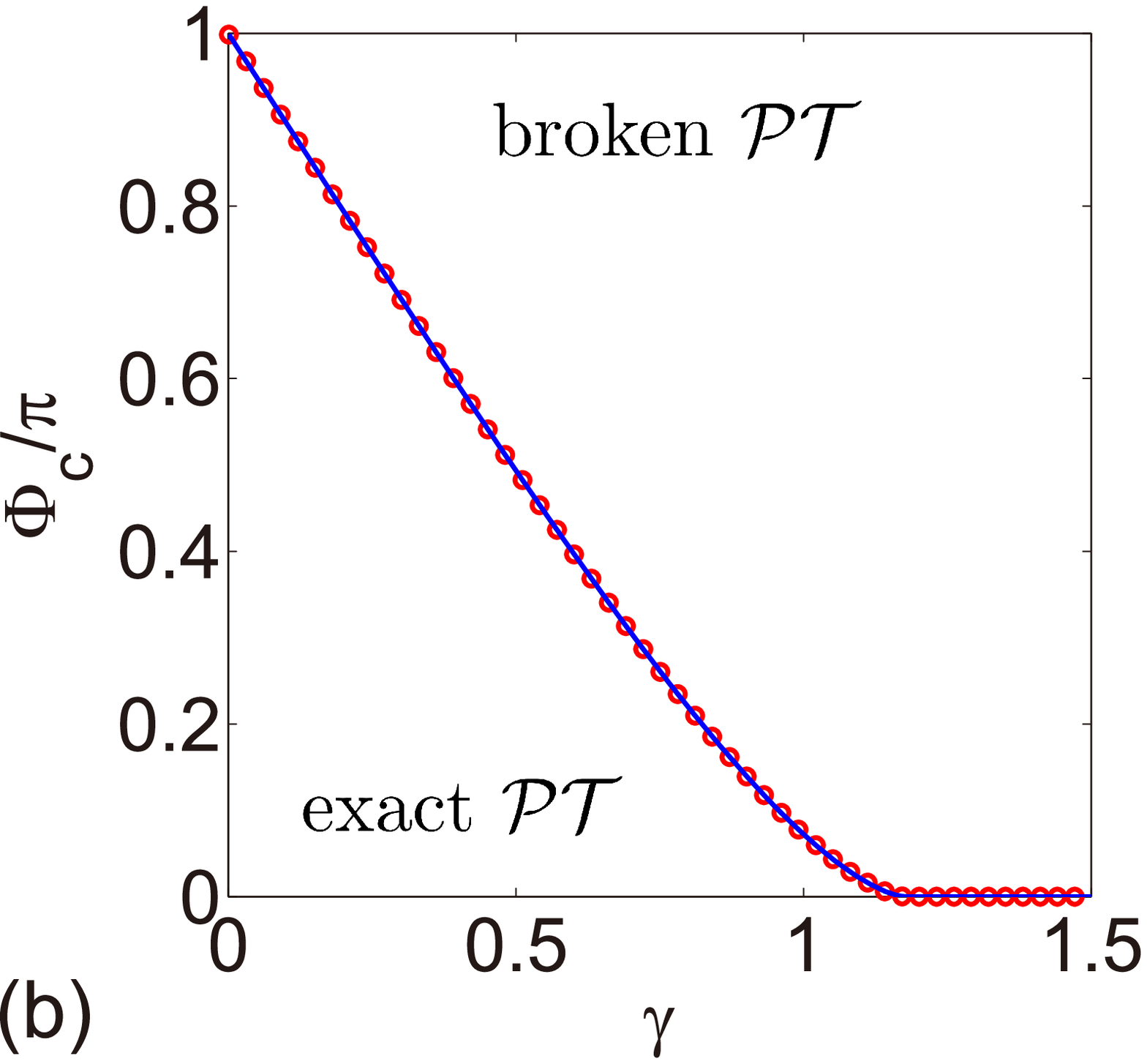}
\caption{(Color online) (a) Numerically determined $k_{\mathrm{c}}$ where $\mathcal{F}(\gamma,0,k)$ is at its
maximum in region $[\pi/2N, \pi/N]$. (b) The numerically determined maximum
magnetic flux $\Phi_{\mathrm{c}}$ as a function of gain/loss $\gamma$. The lower left (upper right) region is the exact (broken) $\mathcal{PT}$-symmetric phase. The ring system is with $N=5$.}
\label{phase_diagram}
\end{figure}

To gain insight regarding $\mathcal{PT}$ symmetry under magnetic flux, we
set the function $\mathcal{F}\left( \gamma ,0,k\right) $ to reach its
maximum at $k_{\mathrm{c}}$ in the region $k\in \lbrack 0,\pi /N]$, where $%
k_{\mathrm{c}}$ can be calculated from \textrm{d}$\mathcal{F}\left( \gamma
,0,k\right) /\mathrm{d}k=0$. We plot $k_{\mathrm{c}}$ and $\Phi _{\mathrm{c}%
} $ in Fig.~\ref{phase_diagram}; the blue lines are obtained by numerically
solving the equation \textrm{d}$\mathcal{F}\left( \gamma ,0,k\right) /%
\mathrm{d}k=0$. The red circles are obtained by numerically diagonalizing
the ring system Hamiltonian; the results from these two methods are
in accord with each other. In Fig.~\ref{phase_diagram}a, we plot $Nk_{%
\mathrm{c}}$ as a function of $\gamma $ for the ring system with $N=5$. As $%
\gamma $ increases from $0 $ to $\gamma _{\mathrm{c}}$, $k_{\mathrm{c}}$
increases from $\pi /(2N)$ to $\pi /N $, the maximal value of $\mathcal{F}%
\left( \gamma ,0,k_{\mathrm{c}}\right) $ decreases from $1$ to $0$, and the
maximum magnetic flux $\Phi _{\mathrm{c}}$ decreases from $\pi $ to $0$
(Fig.~\ref{phase_diagram}b). Figure~\ref{phase_diagram}b shows the $\mathcal{%
PT}$-symmetry phase diagram of the ring system. We can see the sharp changes
at $Nk_{\mathrm{c}}=\pi $ in Fig.~\ref{phase_diagram}a, and $\Phi _{\mathrm{c%
}}=0$ at $\gamma _{\mathrm{c}}\approx 1.1756$ in Fig.~\ref{phase_diagram}b.

\begin{figure}[t]
\includegraphics[ bb=25 20 450 420, width=7.0 cm, clip]{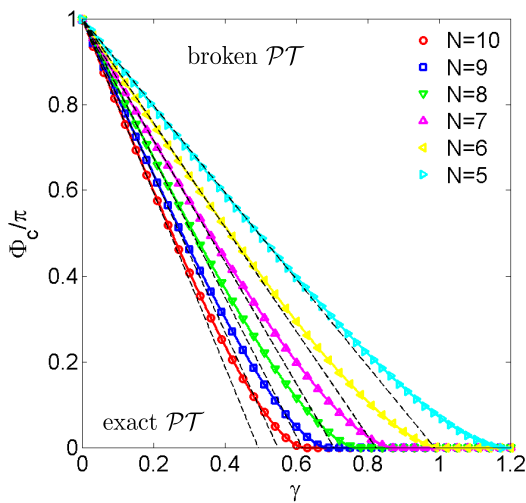}
\caption{(Color online) Maximum
magnetic flux $\Phi_{\mathrm{c}}$ as a function of gain/loss $\gamma$ for a ring system with $N=5$ to $10$.
The color lines (markers) are numerically obtained theoretical analysis (exact diagonalization) of the Hamiltonian $H_{\rm{sc}}$. The dashed black lines indicate $\Phi _{\mathrm{c}}/\pi \approx 1-(2N/\pi
^{2})\gamma $.} \label{Phi_c}
\end{figure}

The maximal magnetic flux ${\Phi }_{\mathrm{c}}$ keeps the ring system in
the exact $\mathcal{PT}$-symmetric phase; it satisfies $\mathcal{F}\left(
\gamma ,\Phi _{\mathrm{c}},k_{\mathrm{c}}\right) =0$. We numerically
calculate the maximum magnetic flux $\Phi _{\mathrm{c}}$, which increases as
the gain/loss $\gamma $ decreases. The maximal magnetic flux $\Phi _{\mathrm{%
c}}$ as a function of $\gamma $ is shown in Fig.~\ref{Phi_c}; the colored
lines and markers are numerical results from the theoretical analysis and
exact diagonalization of $H_{\mathrm{sc}}$, respectively. For weak $\gamma $
($\gamma \lesssim \pi /N$), we approximately have $k_{\mathrm{c}}\approx \pi
/(2N)$. Correspondingly, the maximal magnetic flux is approximately linearly
dependent on the gain/loss $\gamma $ as $\Phi _{\mathrm{c}}/\pi \approx
1-(2N/\pi ^{2})\gamma $ (indicated by dashed black lines in Fig.~\ref{Phi_c}). The
maximum magnetic flux $\Phi _{\mathrm{c}}$ decreases as ring system size $N$
increases.

The $\mathcal{PT}$-symmetric phase transition is closely related to the
magnetic flux in the coupled resonators enclosed magnetic flux. Through
tuning the coupling position of the auxiliary resonator between resonator $1$
and $2N$ (Fig.~\ref{fig1}b), the path lengths of forward- and backward-going
directions change, which linearly affects the magnetic flux. The $\mathcal{PT%
}$ symmetry breaking point varies with the magnetic flux; the gain/loss
threshold for $\mathcal{PT}$ symmetry breaking can be greatly reduced by
increasing the enclosed magnetic flux from zero to half a quantum. The $%
\mathcal{PT}$ symmetry breaking energy levels change from one pair to $N$
pairs when the enclosed magnetic flux is so tuned. The $\mathcal{PT}$
symmetry breaking can be observed at low balanced gain and loss values in
coupled resonators with enclosed magnetic flux.

\section{Conclusion}

\label{V} We investigate $\mathcal{PT}$-symmetric non-Hermitian coupled
resonators in ring configuration threaded by an effective magnetic flux. The
ring system is described by a $2N$-site tight-binding model; the ring system
has a balanced pair of gain and loss located at two opposite sites. We
demonstrate that the system's $\mathcal{PT}$ symmetry is extremely sensitive
to the enclosed magnetic flux when the gain/loss is above a threshold $%
\gamma _{\mathrm{c}}=2\sin (\pi /N)$. We find the minimal number of
conjugate pairs emerging in the system spectrum when the $\mathcal{PT}$
symmetry is breaking in the presence of nontrivial magnetic flux. The system
eigenvalues all become conjugate pairs at a magnetic flux $\Phi =2m\pi +\pi $
($m\in \mathbb{Z} $) for any nonzero gain/loss $\gamma $; or at gain/loss
twice larger than the hopping strength ($\gamma >2$) for any nontrivial
magnetic flux $\Phi $. We show the $\mathcal{PT}$-symmetric phase diagram in
the parameter spaces of $\gamma $ and $\Phi $. The results indicate the
maximal magnetic flux approximately linearly depends on $\gamma $ as $\Phi _{%
\mathrm{c}}/\pi \approx 1-(2N/\pi ^{2})\gamma $ in the weak $\gamma $ region
($\gamma \lesssim \pi/N$). The maximal magnetic flux decreases as the
gain/loss or system size increases. Our findings indicate the $\mathcal{PT}$
symmetry is very sensitive to the nonlocal vector potential in this $%
\mathcal{PT}$-symmetric non-Hermitian system. These results could be useful
in quantum metrology in the future.

\acknowledgments We acknowledge the support of National Natural Science
Foundation of China (CNSF Grant No. 11374163), National Basic Research
Program of China (973 Program Grant No. 2012CB921900), and the Baiqing plan
foundation of Nankai University (Grant No. ZB15006104).


\begin{thebibliography}{99}
\bibitem{ES} W. Ehrenberg and R. E. Siday, Proc. Phys. Soc. (London) \textbf{%
B62}, 8 (1949).

\bibitem{AB} Y. Aharonov and D. Bohm, Phys. Rev. Lett. \textbf{115}, 485
(1959).

\bibitem{MOeffect} K. Fang and S. Fan. Phys. Rev. A \textbf{88}, 043847
(2013).

\bibitem{DynamicModulation} K. Fang, Z. Yu, and S. Fan. Nature Phot. \textbf{%
6}, 782 (2012).

\bibitem{PhotPhot} E. Li, B. J. Eggleton, K. Fang, and S. Fan, Nat. Commun.
\textbf{5}, 3225 (2014).

\bibitem{Hafezi2011} M. Hafezi, E. A. Demler, M. D. Lukin, and J. M. Taylor.
Nature Phys. \textbf{7}, 907 (2011).

\bibitem{Hafezi2013} M. Hafezi, S. Mittal, J. Fan, A. Migdall, and J. M.
Taylor. Nature Phot. \textbf{7}, 1001 (2013).

\bibitem{Chong} G. Q. Liang and Y. D. Chong Phys. Rev. Lett. \textbf{110},
203904 (2013).

\bibitem{Longhi2015OL} S. Longhi, Opt. Lett. \textbf{40}, 1278 (2015).

\bibitem{LXQ} X. Q. Li, X. Z. Zhang, G. Zhang, and Z. Song. Phys. Rev. A
\textbf{91}, 032101 (2015).

\bibitem{Bender98} C. M. Bender and S. Boettcher, Phys. Rev. Lett. \textbf{80%
}, 5243 (1998); C. M. Bender, D. C. Brody, and H. F. Jones, Phys. Rev. Lett.
89, 270401 (2002).

\bibitem{Ali02} A. Mostafazadeh, J. Math. Phys. 43, 205 (2002); J. Math.
Phys. 43, 3944 (2002).

\bibitem{Znojil99} M. Znojil, Phys. Lett. A \textbf{259}, 220 (1999); Phys.
Lett. A \textbf{264}, 108 (1999); Phys. Lett. A \textbf{285}, 7 (2001).

\bibitem{Dorey01} P. Dorey, C. Dunning, and R. Tateo, J. Phys. A \textbf{34}%
, 5679 (2001).

\bibitem{Musslimani OL} R. El-Ganainy, K. G. Makris, D. N. Christodoulides,
and Z. H. Musslimani, Opt. Lett. \textbf{32}, 2632 (2007).

\bibitem{QuanOptAnology} S. Longhi, Laser \& Phot. Rev. \textbf{3}, 243
(2009).

\bibitem{AGuo} A. Guo, G. J. Salamo, D. Duchesne, R.Morandotti, M.
Volatier-Ravat, V. Aimez, G. A. Siviloglou, and D. N. Christodoulides, Phys.
Rev. Lett. \textbf{103}, 093902 (2009).

\bibitem{CERuter} C. E. R\"{u}ter, K. G. Makris, R. El-Ganainy, D. N.
Christodoulides, M. Segev, and D. Kip, Nature Phys. \textbf{6}, 192 (2010).

\bibitem{CPA} Y. D. Chong, L. Ge, H. Cao, and A. D. Stone Phys. Rev. Lett.
\textbf{105}, 053901 (2010).

\bibitem{HChen} Y. Sun, W. Tan, H. Q. Li, J. Li, and H. Chen, Phys. Rev.
Lett. 112, 143903 (2014).

\bibitem{AliPRL2009} A. Mostafazadeh, Phys. Rev. Lett. \textbf{102}, 220402
(2009); Phys. Rev. Lett. \textbf{110}, 260402 (2013).

\bibitem{Lin} Z. Lin, H. Ramezani, T. Eichelkraut, T. Kottos, H. Cao, and D.
N. Christodoulides, Phys. Rev. Lett. \textbf{106}, 213901 (2011).

\bibitem{Feng} L. Feng, Y. L. Xu, W. S. Fegadolli, M. H. Lu, J. E. B.
Oliveira, V. R. Almeida, Y. F. Chen, and A. Scherer, Nature Mater. \textbf{12%
}, 108 (2013).

\bibitem{PengNP} B. Peng, S. K. \"{O}zdemir, F. Lei, F. Monifi, M.
Gianfreda, G. L. Long, S. Fan, F. Nori, C. M. Bender, and L. Yang, Nature
Phys. \textbf{10}, 394 (2014).

\bibitem{PengNC} B. Peng, S. K. \"{O}zdemir, W. Chen, F. Nori, and L. Yang,
Nat. Commun. \textbf{5}, 5082 (2014).

\bibitem{Chang} L. Chang, X. Jiang, S. Hua, C. Yang, J. Wen, L. Jiang, G.
Li, G. Wang, and M. Xiao, Nature Phot. \textbf{8}, 524 (2014).

\bibitem{Jing} H. Jing, S. K. \"{O}zdemir, X.-Y. L\"{u}, J. Zhang, L. Yang,
and F. Nori, Phys. Rev. Lett. \textbf{113}, 053604 (2014).

\bibitem{PengScience} B. Peng, S. K. \"{O}zdemir, S. Rotter, H. Yilmaz, M.
Liertzer, F. Monifi, C. M. Bender, F. Nori, and L. Yang, Science \textbf{346}%
, 328 (2014).

\bibitem{FengZhang} L. Feng, Z. J. Wong, R.-M. Ma, Y. Wang, and X. Zhang,
Science \textbf{346}, 972 (2014).

\bibitem{Hodaei} H. Hodaei, M.-A. Miri, M. Heinrich, D. N. Christodoulides,
and M. Khajavikhan, Science \textbf{346}, 975 (2014).

\bibitem{Bendix} O. Bendix, R. Fleischmann, T. Kottos, and B. Shapiro, Phys.
Rev. Lett. \textbf{103}, 030402 (2009).

\bibitem{Longhi1} S. Longhi, Phys. Rev. Lett. \textbf{103}, 123601 (2009);
Phys. Rev. B \textbf{80}, 235102 (2009); Phys. Rev. B \textbf{81}, 195118
(2010); Phys. Rev. A \textbf{82}, 032111 (2010); Phys. Rev. B \textbf{82},
041106(R) (2010).

\bibitem{Longhi2} S. Longhi, J. Phys. A: Math. Theor. \textbf{47}, 165302
(2014).

\bibitem{Znojil1} M. Znojil, J. Phys. A: Math. Theor. \textbf{41}, 292002
(2008); Phys. Rev. A \textbf{82}, 052113 (2010); J. Phys. A: Math. Theor.
\textbf{44}, 075302 (2011).

\bibitem{Znojil2} M. Znojil, J. Phys. A: Math. Theor. \textbf{47}, 435302
(2014); Ann. Phys. (NY) \textbf{361}, 226 (2015).

\bibitem{Jin} L. Jin and Z. Song, Phys. Rev. A \textbf{80}, 052107 (2009);
Phys. Rev. A \textbf{81}, 032109 (2010).

\bibitem{Joglekar1} Y. N. Joglekar, D. Scott, M. Babbey, and A. Saxena,
Phys. Rev. A \textbf{82}, 030103(R) (2010).

\bibitem{Joglekar2} Y. N. Joglekar and A. Saxena, Phys. Rev. A \textbf{83},
050101(R) (2011).

\bibitem{Joglekar3} D. D. Scott and Y. N. Joglekar, Phys. Rev. A \textbf{83}%
, 050102(R) (2011).

\bibitem{Joglekar4} Y. N. Joglekar, D. D. Scott, and A. Saxena, Phys. Rev. A
\textbf{90}, 032108 (2014).

\bibitem{Joglekar2012} D. D. Scott and Y. N. Joglekar, Phys. Rev. A \textbf{%
85}, 062105 (2012).

\bibitem{JinHu} L. Jin and Z. Song, Phys. Rev. A \textbf{84}, 042116 (2011);
W. H. Hu, L. Jin, Y. Li, and Z. Song, Phys. Rev. A \textbf{86}, 042110
(2012).

\bibitem{Joglekar2013} H. Vemuri and Y. N. Joglekar, Phys. Rev. A \textbf{87}%
, 044101 (2013).

\bibitem{Longhi2013} S. Longhi, Phys. Rev. A \textbf{88}, 062112 (2013).

\bibitem{Hafezi2014} M. Hafezi, Phys. Rev. Lett. \textbf{112}, 210405 (2014).
\end{thebibliography}
\end{document}